\documentclass{article}
\usepackage{emulateapj,epsfig}

\slugcomment{To appear in ApJ April 20 issue, 2004}

\makeatletter

\newenvironment{inlinefigure}{%
\def\@captype{figure}%
\noindent\begin{minipage}{0.999\linewidth}\begin{center}}
{\end{center}\end{minipage}\smallskip}
\makeatother

\def\ltsima{$\; \buildrel < \over \sim \;$}
\def\simlt{\lower.5ex\hbox{\ltsima}}
\def\gtsima{$\; \buildrel > \over \sim \;$}
\def\simgt{\lower.5ex\hbox{\gtsima}}

\begin{document}

\title{The Era of Massive Population~III Stars:
Cosmological Implications and Self-Termination}

\author{Naoki Yoshida\altaffilmark{1}, Volker Bromm, and Lars Hernquist}
\affil{Harvard-Smithsonian Center for Astrophysics, 60 Garden Street,
Cambridge, MA 02138;\\
nyoshida@cfa.harvard.edu, vbromm@cfa.harvard.edu, lars@cfa.harvard.edu}

\altaffiltext{1}{National Astronomical Observatory of Japan,
Mitaka, Tokyo 181-8588, Japan.}

\begin{abstract}
The birth and death of the first generation of stars have important
implications for the thermal state and chemical properties of the
intergalactic medium (IGM) in the early universe.
Sometime after recombination, the neutral, chemically pristine gas was
reionized by ultraviolet photons emitted from the first stars, but also
enriched
with heavy elements when these stars ended their lives as
energetic supernovae.
Using the results from previous high-resolution cosmological simulations of
early structure formation that include radiative transfer,
we show that a significant volume fraction of the IGM
can be metal-polluted, as well as ionized, by massive
Population III stars formed in small-mass ($\sim 10^{6}-10^{7} M_{\odot}$) halos
early on.
If most of the early generation stars die
as pair-instability supernovae with energies up to $\sim 10^{53}$~ergs,
the volume-averaged mean metallicity will quickly
reach $Z\sim 10^{-4}Z_{\odot}$ by a redshift of $\sim 15-20$,
possibly causing a prompt transition to the formation of
a stellar population that is dominated by low-mass stars.
In this scenario, the early chemical enrichment history should
closely trace the reionization history of the IGM, and the end of
the Population~III era is marked by the completion of reionization
and pre-enrichment by $z\sim 15$.
We conclude that, while the pre-enrichment
may partially account for the ``metallicity-floor'' in high-redshift Lyman-$\alpha$
clouds, it does not significantly affect the elemental abundance in the intracluster medium.
\end{abstract}


\section{Introduction}
Chemical elements heavier than lithium are thought to be
produced exclusively through stellar nucleosynthesis.
The primordial cosmic gas remains chemically pristine
until the first supernova (SN) explosions expel metals
that are produced in the precursor stars.
High-redshift observations of the
Lyman-$\alpha$ forest (e.g., Songaila 2001; Pettini et al. 2003),
damped Lyman-$\alpha$ systems (e.g., Prochaska 2002), and
quasars (e.g., Freudling, Corbin, \& Korista 2003)
all indicate that heavy elements are distributed in
various cosmic environments by $z\sim 5$.
While the actual transport mechanism and the nature
of the primary sources have not been determined, it is often
suggested that the origin of
these heavy elements may be attributed to the first generation of stars, the so-called
Population~III stars
(e.g., Ostriker \& Gnedin 1996; Madau, Ferrara, \& Rees
2001; Thacker, Scannapieco, \& Davis 2002; Qian, Sargent, \& Wasserburg 2002).

In the standard cosmological models based on cold dark matter (CDM),
the first cosmological objects are predicted to form
at redshifts greater than 20
(e.g., Couchman \& Rees 1986; Tegmark et al. 1997;
Abel et al. 1998; Yoshida et al. 2003a). Thus,
the first heavy elements are likely to have been processed at such
early epochs. Intriguingly, the recent measurement of a large
Thomson optical depth by the
{\it Wilkinson Microwave Anisotropy Probe (WMAP)}
satellite (Kogut et al. 2003; Spergel et al. 2003)
provides evidence that the universe was
reionized very early on,
supporting the above notion that the first stars could
form at $z\ga 20$ (e.g., Cen 2003;
Haiman \& Holder 2003; Wyithe \& Loeb 2003a,b;
Sokasian et al. 2003a).

 The relative abundances of heavy elements
produced in the first stars are of great importance because
observations of the elemental abundance pattern of ultra metal-poor stars (e.g.,
Christlieb et al. 2002) can place  strong constraints
on the environments in which these stars were formed
and possibly on the progenitor mass (Umeda \& Nomoto 2003; Schneider et al. 2003a).
Theoretical modeling of the formation of the first stars
(Abel, Bryan, \& Norman 2002; Bromm, Coppi, \& Larson 2002;
Omukai \& Palla 2003)
consistently indicate that they
were rather massive, with
characteristic mass being possibly several hundred solar masses.
If the first stars are indeed as massive as $\sim 200 M_{\odot}$,
they end their lives as energetic SNe via
the pair-instability mechanism (e.g., Barkat, Rakavy, \& Sack 1967;
Bond, Arnett, \& Carr 1984; Fryer, Woosley, \& Heger 2001; Heger \& Woosley 2002;
Woosley, Heger, \& Weaver 2002), releasing a total energy of up to
$\sim 10^{53}$~ergs.
Such energetic explosions in the early universe
are violently destructive:
they expel the ambient gas out
of the gravitational potential well of small-mass dark matter
halos, causing an almost complete evacuation
(Bromm, Yoshida, \& Hernquist 2003, hereafter Paper~I;
Wada \& Venkatesan 2003).

Since the massive stars process a substantial fraction of
their mass into heavy elements,
early SN explosions may provide
an efficient mechanism to pollute the surrounding intergalactic medium (IGM).
Models of stellar evolution (e.g., Heger \& Woosley 2002)
predict that, in massive zero-metallicity stars,
up to $\sim 40-50$\% of the total stellar mass
is processed into heavy elements in the core
and,
in the pair-instability case,
{\it all} of it is finally ejected during the explosion
without leaving any compact remnant behind.
Blast waves triggered by the SN explosions will
then drive supersonic flows of gas that are enriched with the
processed heavy elements.
Consequently, because of the
short lifetimes of such massive stars, the emergence of
the early generation stars should be almost immediately
followed by metal enrichment of the IGM; i.e.,
prompt metal enrichment could be achieved
efficiently by Population~III SNe in the early universe.

The transport of metals by SN-driven winds has been studied
extensively in the context of conventional galaxy formation
(e.g., Larson 1974; Dekel \& Silk 1986; Vader 1986; Mac Low \& Ferrara 1999;
Aguirre et al. 2001a,b;
Madau et al. 2001; Scannapieco, Thacker, \& Broadhurst 2001; Mori,
Ferrara, \& Madau 2002; Furlanetto \& Loeb 2003; Springel \& Hernquist
2003a,b).
In fact, gas flows induced by SN explosions appear to
be the most plausible way of distributing metals,
which are observed to exist under a broad range of environments.
Interestingly, an array of observations indicate that
the metals were produced and dispersed at early stages of
structure formation.
Spectra of distant quasars consistently show that there
are at least some metals in the diffuse
IGM at $z\simeq 2-3$ (e.g., Schaye et al. 2000, 2003).
Qian \& Wasserburg (2001, 2002) argue that the observed abundance patterns
of metal-poor stars can be explained by the ``prompt-inventory''
of heavy elements from very massive stars.
Loewenstein (2001) points out that a contribution from very
massive metal-free stars may account for the observed abundance
anomalies of oxygen, silicon, and iron in the intracluster medium (ICM).
More recently, Freudling et al. (2003) discovered a strong {\ion{Fe}{2}} emission
feature from a $z\simeq 6$ quasar, providing strong evidence that
iron is produced in some stars/galaxies already at $z\ga 10$,
which is expected if black hole growth is regulated by
star formation (e.g., Di Matteo et al. 2003a,b).
Together with the implication of an early reionization derived from the
{\it WMAP} results, these observations indicate that Population~III stars
may have been an important contributor of heavy elements in the diffuse IGM/ICM as well as
the source of radiation responsible for the 
early reionization of the universe.
The association of the IGM metal enrichment with the UV-photon production
from the first stars was previously discussed by Oh et al. (2001), and more recently
by Stiavelli, Fall, \& Panagia (2003).
In these studies, however, the analysis was not based on a detailed modeling
of the high-redshift star formation rate,
and the precise epochs of Population~III pre-enrichment and reionization therefore
remained somewhat uncertain.

In the present paper, we revisit the cosmic metal production
in the context of early structure formation. We explore the
possibility that the first stars are responsible for reionization
of the universe at $z\ga 15$, while also
causing a significant, prompt chemical enrichment of the IGM.
We argue that, if the early reionization was the
result of very massive
Population~III stars, the mean metallicity should reach a critical level
at which the efficiency of gas cooling is greatly enhanced, thus
changing the overall mode of star formation (e.g., Mackey, Bromm, \&
Hernquist 2003).

The hydrodynamic transport of heavy elements induced by SN explosions
is studied numerically in Paper~I,
using high-resolution, three-dimensional cosmological simulations.
We show that a large amount of the gas
is expelled out of the shallow gravitational
potential of the host ``minihalos''.
The simulations in Paper~I further demonstrate that,
if the explosion kinetic energy is as high as $\sim 10^{53}$~ergs,
nearly 90\% of the pre-enriched gas in the immediate vicinity
of the explosion site eventually escapes out of the host
halo, reaching up to a distance of $\sim 0.5$~kpc in a few million years.
Since multiple SNe are expected to be occurring at $z\ga 15$
in distant star-forming regions within a cosmological volume,
the total amount of metals produced {\it and} distributed
into the IGM may well exceed the critical level required to
affect the thermal and chemical properties of the IGM.
The transition in stellar populations, from Population~III to Population~II,
may then be naturally caused early on.

Throughout the present paper, except in section \ref{rsi} 
where we consider a variant CDM model,
we work with a flat $\Lambda$-dominated Cold Dark Matter universe
with matter density $\Omega_{\rm m}=0.3$, cosmological constant
$\Omega_{\Lambda}=0.7$ and the Hubble constant at the present time
$h=0.7$ in units of $100$km s$^{-1}$Mpc$^{-1}$.  We set the baryon
density to $\Omega_{\rm b}=0.04$ and the initial density fluctuations
are normalized to $\sigma_8 =0.9$. These parameters
are consistent with those employed in the numerical simulations
we refer to in the following sections.

\section{Feedback from early generation stars}
In this section we summarize various feedback effects
from the first stars, and derive useful order-of-magnitude
estimates for quantities needed to study these effects.
The ultimate fate of a metal-free star
depends critically on its mass. Theoretical stellar models predict
the following classification (Heger \& Woosley 2002; Heger et al. 2003):
(1) $8 < M_{*} < 25 M_{\odot}$: these stars explode as core-collapse SNe and
leave a neutron star behind,
(2) $25 < M_{*} < 40 M_{\odot}$: these explode as faint Type~II SNe and
leave black holes behind after fall-back of most of the envelope
(see Umeda \& Nomoto 2002, 2003),
(3) $40 < M_{*} < 140 M_{\odot}$: these do not explode as SNe and
directly collapse into black holes,
(4)  $100 < M_{*} < 140 M_{\odot}$: these experience a pulsational
instability and eject their outer envelope, again leaving a black hole behind,
(5)  $140 < M_{*} < 260 M_{\odot}$: these explode as pair-instability
SNe, causing complete disruption,
and (6) $M_{*} > 260 M_{\odot}$: these collapse, in the absense of rotation,
directly into massive black holes.

In summary, only the pair-instability case can substantially contribute
to the metal enrichment of the IGM, whereas remnant black holes (or neutron stars) are
left in the other cases, which we shall discuss in \S~2.5.
While early generation stars may be born with a broad range of
masses (e.g., Bromm, Coppi, \& Larson 1999;
Nakamura \& Umemura 2001, 2002; Omukai \& Yoshii 2003),
very massive stars are dominant in causing
various feedback effects to the IGM in the early universe
such as ionization, heating, and chemical enrichment.
Thus, throughout the present paper, we primarily consider very massive
($\sim 100-300 M_{\odot}$) metal-free stars as Population~III stars.

\subsection{Population~III Stars as Reionization Sources}

We first discuss radiative feedback.
The efficiency of ionizing photon production from massive Population~III
stars is obtained from detailed calculations (e.g.,
Bromm, Kudritzki, \& Loeb 2001b; Schaerer 2002, 2003).
As our fiducial case, we consider a metal-free star with mass
$M_{*}=200M_{\odot}$. Such a star emits about $3\times 10^{50}$
ionizing photons per second, and has a main-sequence lifetime
of $\sim 2$ million years.

Consider a population of such massive stars formed in a cubic
comoving-Mpc volume by $z$.
We introduce a quantity
\begin{equation}
q(z) \equiv \frac{N_{\rm ph}(z)}
{N_{\rm H}},
\end{equation}
where $N_{\rm ph}(z)$ is the cumulative number of ionizing photons
emitted from the stars, and $N_{\rm H}=4\times 10^{66}$ is the number of
hydrogen atoms in the volume. This `number of photons per atom'
$q(z)$ serves as an indicator of the global ionization fraction
as a function of $z$.
Clearly, in order to completely ionize the IGM, at least the same number
of UV photons as the number of hydrogen atoms must be emitted within the
volume.  It follows from these numbers that more than a few hundred
very massive stars must be formed in a
comoving-Mpc volume to cause complete
reionization.

\subsection{Population III Stars as Sources of Heavy Elements}

Massive metal-free stars are not only luminous ionizing sources,
but also efficient factories of heavy elements.
For stars with masses $\sim 100-300 M_{\odot}$,
typically up to $\sim 50$\% of the mass
is processed into heavy elements in the core,
consisting mostly of carbon, oxygen,
silicon and iron.
We denote this mass fraction with $f_{Z}$, and set it to be 0.5
as our fiducial value for massive stars in the remainder of the
paper. We further introduce a parameter
\begin{equation}
\zeta = \frac{N_{\rm n}}{N_{\gamma}} \mbox{\ ,}
\end{equation}
where $N_{\rm n}$ is the total number of nuclei
in the heavy elements processed in the core,
and $N_{\gamma}$ is the total number of ionizing photons emitted
by the star during its lifetime.
Namely, with $m_{\rm p}$ denoting the proton mass, $m_{\rm p}\times N_{\rm n}$
is the total mass of metals produced in the star.
This quantity gives the ratio of the stellar metal to photon
production which is, in principle, uniquely determined by the
structure and evolution of the respective star.
For a metal-free star with mass $200 M_{\odot}$ that is destined
to die in a pair-instability supernova, the actual value is
approximately
\begin{equation}
\zeta_{\rm PISN} = \frac{f_{Z}\times 200M_{\odot}/m_{p}}{\rm 2\times 10^{64}}
= \frac{10^{59}}{\rm 2\times 10^{64}}\sim 5\times 10^{-6},
\label{eq:zeta}
\end{equation}
for $f_{Z} = 0.5$.
It is straightforward to compute the value for $\zeta$ for less massive
stars
or for stars with finite metallicity
from stellar evolution models.

Intriguingly, this parameter $\zeta_{\rm PISN}$ for the very massive stars
that are destined to explode as pair-instability SNe
can be expressed in terms
of fundamental physical constants as follows. We estimate the total number
of ionizing photons produced during the lifetime of a Population~III star with mass
$M_{\ast}$ as $N_{\gamma}\simeq 0.007 M_{\ast} c^2 / \epsilon_{\rm ion}$.
Here, $\epsilon_{\rm ion}\simeq e^2/a_0\simeq \alpha^2 m_e c^2$ is the
binding energy of an electron in an H atom, with $a_0$ being the Bohr
radius, $\alpha\simeq 1/137$ the fine-structure constant, and $m_e$ the
electron rest mass. Assuming the limiting
case of a 100\% nuclear burning efficiency, we can write for the total
number of nucleons processed into heavy elements: $N_n\simeq M_{\ast}/m_p$.
We now have, for very massive stars that die as 
pair-instability SNe,
\begin{equation}
\zeta_{\rm PISN} = \frac{\alpha^2 m_e}{\alpha_{\rm pn}^2 m_p}\sim 4\times 10^{-6}\mbox{\ ,}
\end{equation}
where $\alpha_{\rm pn}^2\simeq 0.007$ describes the strength of
the pion-nucleon interaction. The production of ionizing photons
and metals in a massive star is thus determined by the ratio of the atomic
and nuclear binding energies.

\subsection{Population III Supernova Explosions}

The processed metals can be distributed into the IGM
only if some transport mechanism operates,
otherwise the metals would be confined within a small region
and the diffuse IGM would be largely unaffected.
In this section, we first describe the hydrodynamics
of a single Population III SN explosion in the early universe
and then show that the gas and the heavy elements
can be transported over a large distance.

Assuming spherical symmetry of the system,
we apply the usual Sedov-Taylor self-similar analysis
for a given explosion energy and density of the ambient medium.
We first need to specify the density profile of the surrounding IGM
in order to fully describe the evolution of the SN blast wave.
For a density profile approximated by a simple power-law,
$\rho \propto r^{-w}$, the expansion of a blast wave is described by
the self-similar solution:
\begin{equation}
R_{\rm s} =\xi \left(\frac{E\;t^2}{n}\right)^{1/(5-w)},
\label{eq:sedov_p}
\end{equation}
where $\xi$ is a constant determined by the energy integral,
$E$ is the explosion energy, $n$ the particle number density
of the surrounding medium, and $t$ the time elapsed since the explosion
(Sedov 1959; see also Ostriker \& McKee 1988).
In reality, the gas density profile can be more complex because
photo-evaporation
due to radiation from the central star is very effective in small halos
($T_{\rm vir} \la 10^4$ K).
Large pressure gradients in the ionized region
drive an outward gas flow {\it before} the explosion is triggered
(see Paper~I).
Hence, although initially the gas density is high around the star,
it is expected to be substantially reduced by the time when
the star explodes,
and the density profile will be much shallower
except in the outer edge where a supersonic shockwave is propagating. Taking
the effective photo-evaporation into account, we assume the density of the
ambient medium around the explosion site
is approximately constant at the mean IGM density at $z$,
\begin{equation}
n_{\rm H, IGM}\simeq 0.0014\left(\frac{\Omega_{\rm b}h^2}{0.02}\right)\left(\frac{1+z}{20}\right)^3,
\end{equation}
where $\Omega_{\rm b}$ is the baryon density and $h$ is the Hubble constant in units of
100 km s$^{-1}$ Mpc$^{-1}$.
Note that the shock radius in equation
(\ref{eq:sedov_p}) scales only weakly with the density and the explosion energy
unless the density profile is very steep.
The three-dimensional simulations in Paper I show that the evolution of the
shock radius is well described by equation (\ref{eq:sedov_p})
in a piecewise manner for regions
having density profiles with different slopes.
For a constant density medium,
the Sedov-Taylor self-similar solution reduces to
\begin{equation}
R_{\rm s}\simeq 200\;\left(\frac{E_{51}}{n_0}\right)^{1/5} t_{7}^{2/5} \;\; {\rm pc},
\label{eq:sedov}
\end{equation}
where $R_{\rm s}$ is the shock radius, $E_{51}$ the total
SN energy in units of 10$^{51}$ ergs, $n_0$ the particle
number density of the ambient medium in units of 1 ${\rm cm}^{-3}$,
and $t_{7}$ the elapsed time since the explosion in units of $10^7$ years.
For the explosion energy,
Heger \& Woosley (2002) calculate the energy released by a Population~III
pair-instability SN to be in the range $\sim 10^{51}-10^{53}$~ergs.

The above solution is valid if the gas remains adiabatic, but
breaks down when the shock becomes radiative.
The hot gas in the interior of high redshift ($z\ga 10$) SN
remnants (SNRs) cools partly via inverse Compton cooling.
The characteristic cooling time for this process is given by
\begin{equation}
t_{\rm Comp} \simeq 10^7 \left(\frac{1+z}{20}\right)^{-4} \; {\rm yr}.
\end{equation}
Note the strong dependence on redshift.
At $z=20$, the SNR will remain approximately adiabatic
(hence equation [\ref{eq:sedov}] is valid) until $\sim 10^7$ years
after the explosion.
However, this provides only a {\it maximum} timescale.
In the post-shock region,
the gas density and temperature are so high that
hydrogen and helium atomic line cooling
is the dominant cooling process.
Thus, a thin dense shell could be formed faster than
the above timescale.  The exact time when the post-shock region becomes
radiative depends on the total explosion energy and the density of the
ambient medium in a complicated manner. The shell forms
when the local cooling time $t_{\rm cool}(r)$ at radius $r$
becomes shorter than the expansion timescale $R_{\rm s}/\dot{R_{\rm s}}$.
We use the analytic Sedov-Taylor solution for the temperature $T(r)$
and density $\rho(r)$ of the gas to compute the local gas cooling rate
$\Lambda (\rho(r), T(r))$,
assuming an explosion energy $E=10^{53}$ ergs
and the gas density $n=0.0014$ cm$^{-3}$ ($\sim$ the mean IGM density
at $z\sim 20$). Recent numerical simulations by Whalen, Abel, \& Norman (2003)
indeed show that photo-evaporation by the central star reduces the final gas density
within the host halo to very small values, comparable to the mean IGM density.
In this case, the shell forms at a time $\sim 3\times 10^6$~yr.
Then the shell radius (the ``stalling-radius'') is $\sim 1.2$ kpc, and it expands only slowly during
the later snow-plow stage as $R_{\rm s} \propto t^{1/4}$
(see Figure 1 in Paper I).
This is perhaps an optimistic estimate because the gas evacuation efficiency
by radiation from the stars depends on the host halo mass and the gas density profile.
For a ten times higher density, $n=0.014$ cm$^{-3}$,
the stalling-radius is calculated to be only 40\% smaller, $\sim 0.7$ kpc, which
is more consistent with the result of our simulation presented in Paper I.
Overall, because of the weak dependence of the self-similar solution (equation [\ref{eq:sedov}])
on the density, the above estimates will not be significantly affected by (possible) variations in the
gas density around the SN sites.

The momentum-driven slow expansion is still important over a cosmological timescale
(Bertschinger 1983), with the radius doubling every twenty folds
in time. Thus, if mixing of the original ejecta with the ambient medium is efficient,
the expelled metals can travel over a large distance ($\sim$ a few kpc)
within a Hubble time.
Three-dimensional numerical simulations of Nakasato \& Shigeyama (2000)
show that the SN ejecta are fairly well mixed within the remnant,
with the enriched volume fraction being close to unity. Wada \& Venkatesan (2003)
show that metal-dispersal is very efficient even in dense, disk-shaped gas clouds.
Therefore, the processed metals are likely to be transported as far
as the shell radius.

\subsection{Global Quantities}

 Let us calculate the evolution of the volume-averaged mean
metallicity of the IGM.
We allow the uncertainty associated with the characteristic mass,
or initial mass function (IMF), of
Population III stars by assuming that a fraction, $f_{\rm PISN}$, of
the Population III stars have masses in the pair-instability range.
Suppose that $n_{*}(z)$ stars are formed by $z$
in a unit comoving volume. The total mass of metals produced within the volume is
then
\begin{equation}
M_{Z}=f_{\rm PISN}\times n_{*}(z) f_{Z}M_{*}.
\end{equation}
We further normalize this quantity to the mass of the IGM within the volume,
to obtain the mean metallicity:
\begin{eqnarray}
\bar{Z}&=&\frac{M_{Z}}{M_{\rm IGM}}
=\frac{n_{*}(z)\times f_{\rm PISN} \times f_{Z}M_{*}}{m_{\rm p}\times N_{\rm H}} \nonumber\\
&\approx&f_{\rm PISN}\times q(z)\times \zeta_{\rm PISN}.
\label{eq:zmean}
\end{eqnarray}
Recall that $\zeta_{\rm PISN}$ is approximately a constant for massive 
metal-free stars that explode as pair-instability SNe
(equation [\ref{eq:zeta}]),
and $q(z)$ traces the IGM ionization fraction.
The above equation simply shows that the total quantity of heavy elements
ejected into the IGM is proportional (weighted by the fraction
$f_{\rm PISN}$) to the number of UV photons emitted by $z$.

We now consider the simplest case where {\it all} of the Population III stars have a mass
of $200 M_{\odot}$
and explode as pair-instability SN, i.e., $f_{\rm PISN}$=1.
We further assume that $q_{\rm reion}\sim 2$ marks the epoch of 
reionization (see, e.g., fig. 8 of Sokasian et al. 2003b).
Setting $\zeta_{\rm PISN}=5\times 10^{-6}$, we obtain
\begin{equation}
\bar{Z}|_{z_{\rm reion}}= 10^{-5}\approx 5\times 10^{-4}Z_{\odot},
\end{equation}
which coincides with the critical metallicity
necessary for the formation of low-mass gas clumps
(Bromm et al. 2001a; see also Mackey et al. 2003).
Note also that very massive stars become progressively unstable
for increasing metallicity (Baraffe, Heger, \& Woosley 2001).
Therefore, for $f_{\rm PISN}=1$, the epoch when the transition in the mode of
star formation occurs should closely match the {\it first} reionization epoch.
Note, however, that we have neglected recombination
in the above argument. If Population~III star formation
occurs rather gradually, taking a considerably longer time
than a recombination time, most of the UV photons are simply consumed
by recombination without increasing the global ionization
fraction of the IGM. Approximating the IGM density with the cosmic mean
density,
we estimate the recombination timescale for hydrogen:
\begin{equation}
t_{\rm rec}\approx 10^{8}{\rm \,yr \,}\left(\frac{1+z}{20}\right)^{-3}.
\end{equation}
If the increase in the number of UV photons is sufficiently
rapid, compared to $t_{\rm rec}$, then the ionization fraction
of the IGM should monotonically increase. This is indeed the case for
the model we describe below.

 An important question is {\it when and how rapidly} did the early generation
stars emerge? 
We address this issue using the formation rate of Population III
stars derived by Yoshida et al. (2003a). Their result is based on a
semi-analytic
model coupled with the outputs of a large cosmological $N$-body simulation.
In their model, star-forming regions are identified
as halos in which the gas has been able to cool via molecular hydrogen
cooling,
and the usual `one-star-per-halo' assumption is made
such that only a single star is formed per star-forming gas cloud.
The model also includes the negative feedback owing to photo-dissociation
of molecular hydrogen by photons in the Lyman-Werner band.
Hence, the model provides a conservative estimate for the early
star-formation rate (SFR). It is worth noting that the resulting
SFR was found to be in good agreement with that
obtained from the hydrodynamic simulations
of Sokasian et al. (2003a) which include radiative transfer
as well as various feedback processes.
We revised the empirical fit to the SFR in Yoshida et al. (2003a)
and found that it is well-described by a particular form proposed by
Hernquist \& Springel (2003):
\begin{equation}
\dot{\rho}_{*}(z)=\dot{\rho}_{*, 0}\frac{\chi^{p}}{1+\tilde{\alpha}
(\chi-1)^{\tilde{q}} \exp(\beta \chi^{r})},
\label{eq:sfr_popIII}
\end{equation}
where
\begin{equation}
\chi (z)\equiv \left(\frac{H(z)}{H_0}\right)^{2/3}
\end{equation}
and, in this case,
$p=5$, $\tilde{q}=1$, $r=1.85$, $\tilde{\alpha}=0.01$, $\beta=0.04$,
$\dot{\rho}_{*, 0}=3.0\times 10^{-8} M_\odot$ yr$^{-1}$ Mpc$^{-3}$.

Figure 1 shows the Population III SFR. The flattening at $z<25$ reflects the
suppression of the gas cooling due to global radiative feedback,
showing the self-regulating nature of early star formation
(e.g., Ricotti, Gnedin, \& Shull 2002; Wise \& Abel 2003; Yoshida et al. 2003a).
We compare the derived SFR with
the result of the hydrodynamic simulations of Sokasian et al. (2003a).
To this end, we have
averaged over all the nine models considered in Sokasian et al. (2003a)
which differ in the star-formation efficiency as well as the strengths
of the feedback effects.
Overall, the agreement between the semi-analytic
prediction and the simulation results is quite good.
As Sokasian et al. (2003a) note,
the SFR derived from their simulations may suffer from
finite boxsize effects at high redshifts ($z>25$), which
accounts for the small discrepancy between the analytic model and
the simulation result seen in Figure 1 (see also Yoshida et al. 2003b;
Barkana \& Loeb 2003)

In the middle panel of Figure 1, we plot the cumulative number of
ionizing photons per baryon $q(z)$.
We compute the number of ionizing photons
assuming a single massive ($M_{*}=200M_{\odot}$) star emits
ionizing photons at a rate of $3\times 10^{50}$ s$^{-1}$
during its main-sequence lifetime (Bromm et al. 2001b).
The dashed line in the middle panel indicates $q=1$,
which is achieved by $z=17$.
We note that the total number of UV photons increases by more
than an order of magnitude between $z=22$ and $z=17$.
The corresponding time interval
is $\sim 50$ Myr, whereas the recombination timescale at $z\sim 20$
is $\sim 100$ Myr. Therefore, to lowest order, the photon consumption
due to hydrogen recombination can be neglected.
In addition, since the photon escape fraction from
minihalos is expected
to be of order unity (Whalen et al. 2003),
and the clumping of the IGM at these early epochs is small
(Haiman, Abel \& Madau 1999), $q\sim$ a few should
suffice to reionize a substantial fraction of the IGM
(e.g., Sokasian et al. 2003b).
It is worth mentioning that, in our model,
the time when $q=1$ coincides with the epoch of reionization
suggested by the {\it WMAP} data, $z=17\pm4$ (assuming an instantaneous
reionization, see Kogut et al. 2003).
At the same time, as shown in the bottom panel in Figure 1,
the global mean metallicity reaches the metallicity floor
of a few $\times 10^{-4}Z_{\odot}$.
We emphasize that this {\it global} mean is merely an
approximate estimate because,
locally around the SN sites where the formation
of second generation stars is likely to take place,
the gas metallicity could be much higher than this value.

\begin{inlinefigure}
\resizebox{8.8cm}{!}{\includegraphics{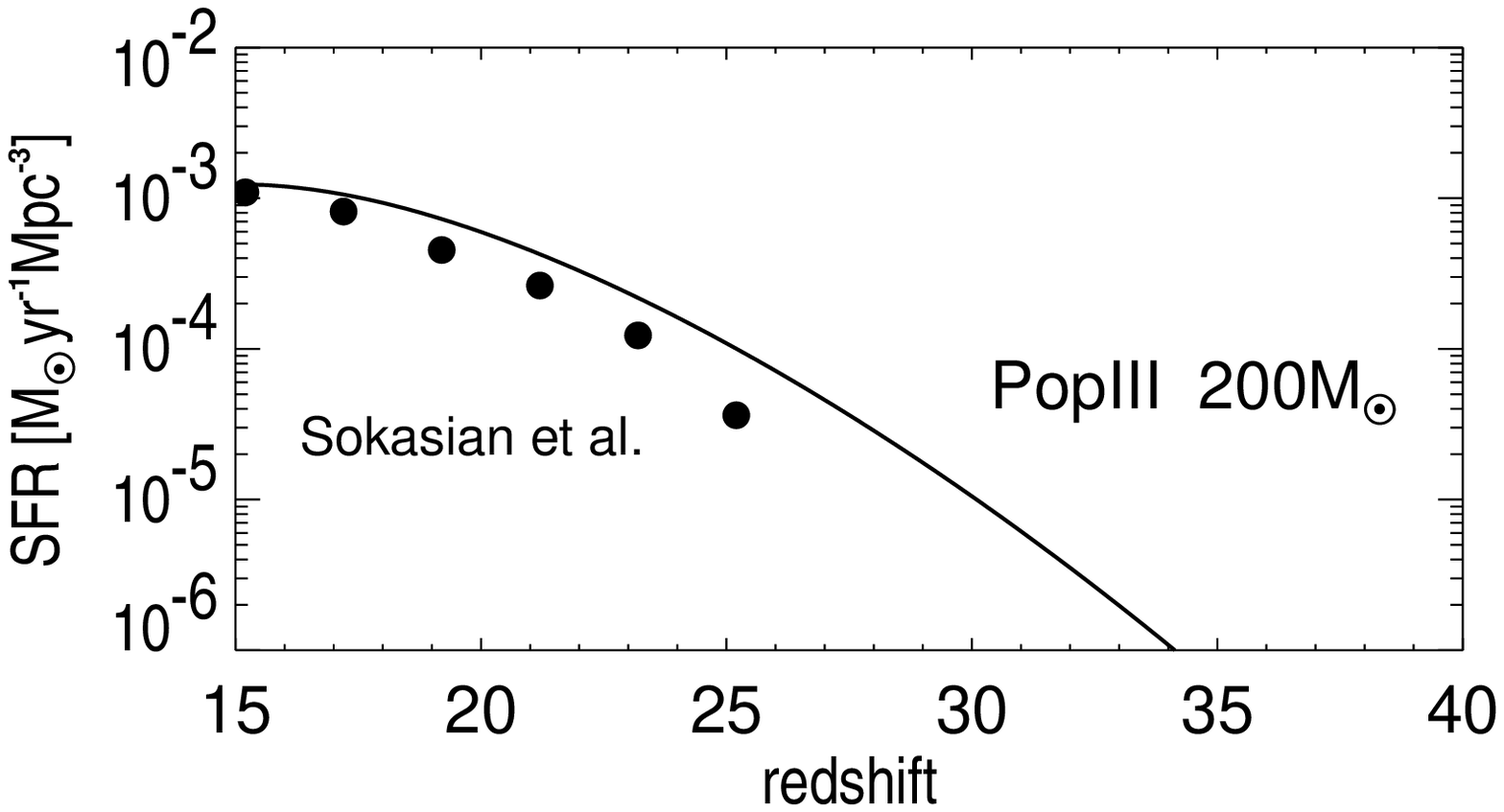}}
\vspace{-3mm}
\resizebox{8.8cm}{!}{\includegraphics{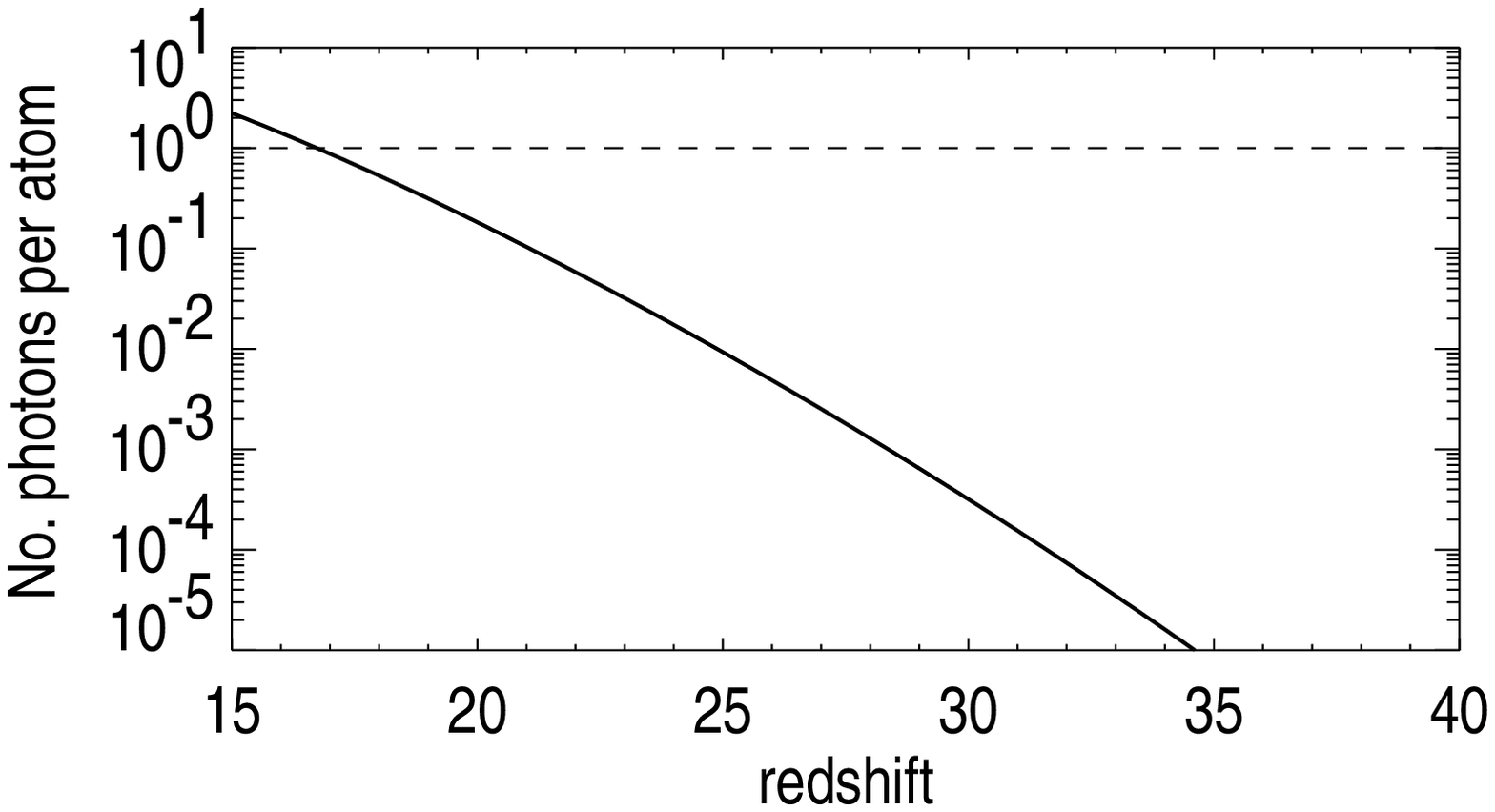}}
\vspace{-3mm}
\resizebox{8.8cm}{!}{\includegraphics{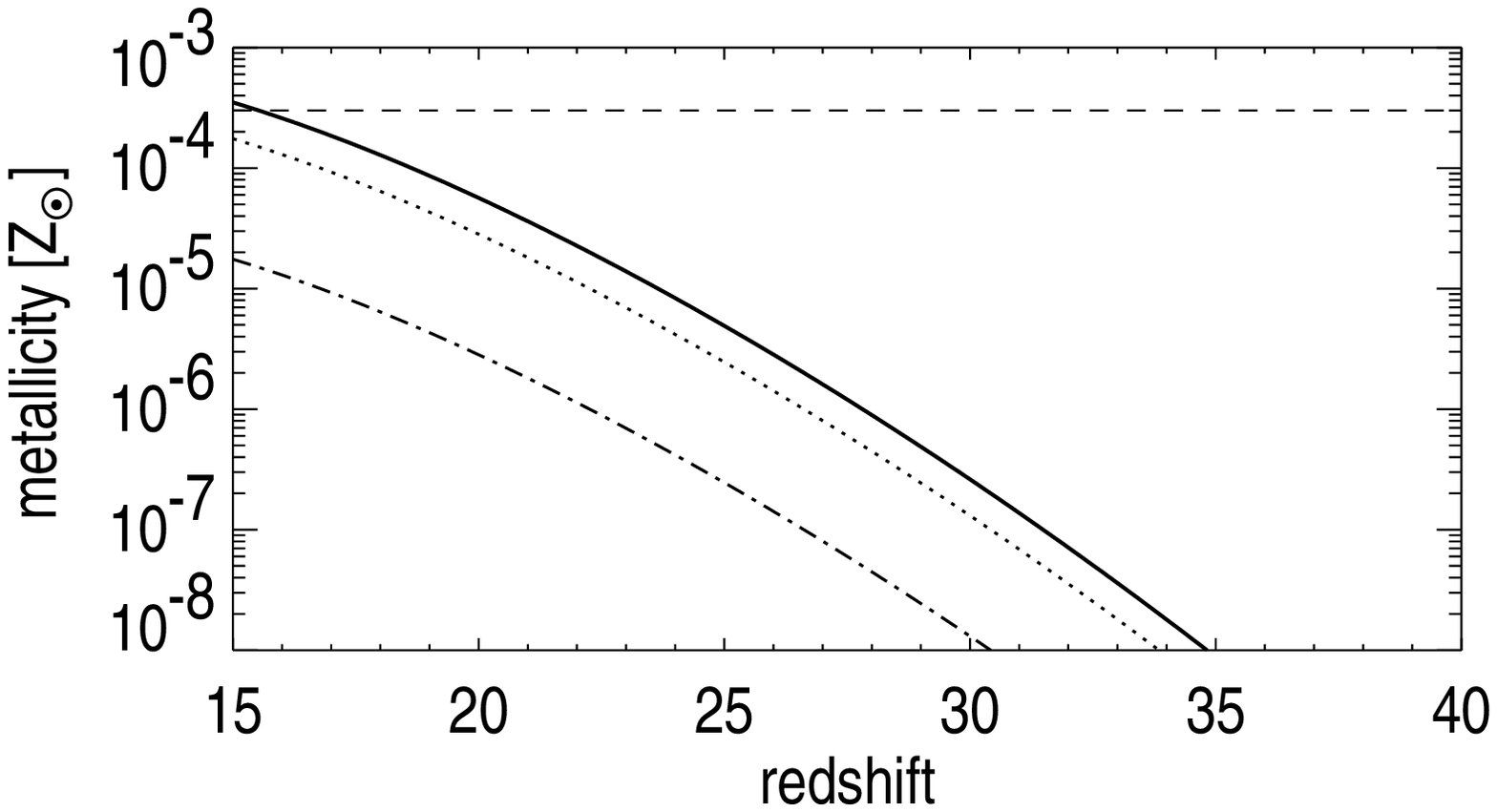}}
\caption{The global star formation
rate density (top), 
the cumulative number of photons per hydrogen atom (middle),
and the mean IGM metallicity $\bar{Z}$ (bottom), as functions of
redshift.
In the top panel we compare our semi-analytic prediction
(solid line) with the result of numerical simulations of Sokasian
et al. (2003a) (filled circles).
The dashed line in the middle panel indicates $q=1$.
The critical metallicity $10^{-3.5} Z_{\odot}$ is also
indicated by the dashed line in the bottom panel.
We plot three cases with mixing
efficiency of the heavy elements with the diffuse IGM,
100\% (solid line), 50\% (dotted line), and 5\% (dot-dashed line).
\label{plot1}}
\end{inlinefigure}

\subsection{Remnant Black Holes}

In the previous sections, we have exclusively considered
very massive
stars that explode as pair-instability SNe, i.e., those with masses
$140 - 260 M_{\odot}$.
Stars with masses outside this range are likely to leave remnant
black holes behind (Fryer et al. 2001).
In particular, stars even more massive than $260 M_{\odot}$ completely collapse
to black holes, which could be the seeds of supermassive black holes
(Madau \& Rees 2001; Islam, Taylor, \& Silk 2003;
Volonteri, Haardt, \& Madau 2003).

It is straightforward to compute the evolution of the
mass density locked up in black holes using our model Population III SFR.
Let us adopt the simplest model
where all the Population III stars are extremely massive and
become black holes.
Although this is an implausible situation because
then the IGM would remain chemically pristine forever
and the ``star-formation conundrum'' would arise
(see Schneider et al. 2002),
it is illustrative to study the evolution of the black hole mass density
in this scenario.

We consider three cases with $M_{*}=300, 600, 1000 M_{\odot}$,
and the resulting mass density is plotted in Figure \ref{plot2}.
Note that our model for the Population III star formation
is based on the `one-star-per-halo' assumption
and thus the star formation rate density scales approximately
with $M_{*}$.
Figure \ref{plot2} shows that the mass density at $z\sim 15$ is comparable to
the present-day mass density in black holes in the local universe
(e.g., Merritt \& Ferrarese 2001) for high mass values.
Note that these Population III blackholes are likely to have sunk
to the galactic centers by the present epoch due to dynamical friction
(e.g., Madau \& Rees 2001).

Obviously, this extreme top-heavy mode of Population III
star formation should be stopped at some point.
Otherwise too much baryonic mass would be converted
into black holes locally by $z=0$.

\begin{inlinefigure}
\resizebox{8cm}{!}{\includegraphics{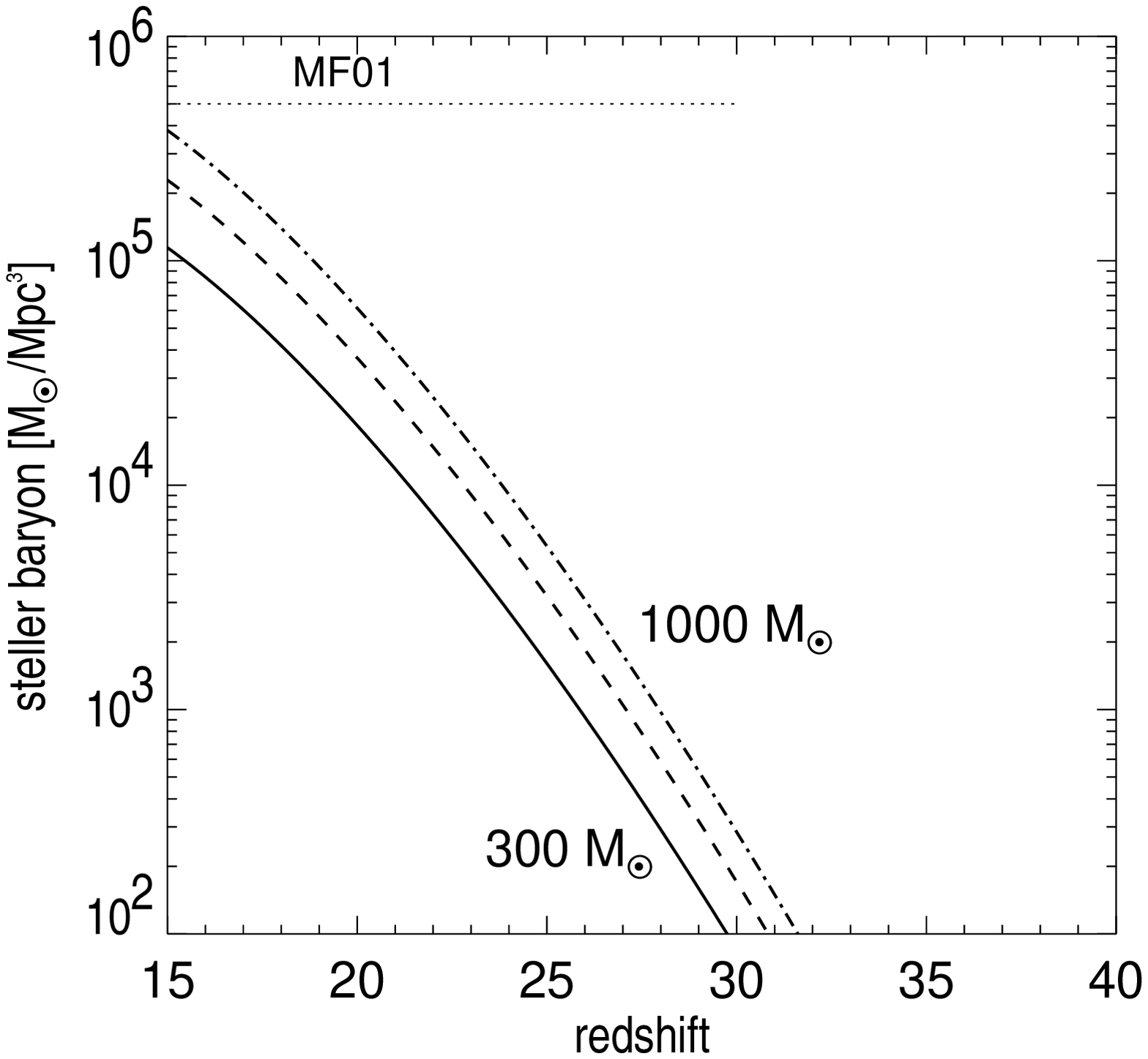}}
\caption{The total mass of baryons converted to stars per unit comoving volume,
for $M_{*}=300$ (solid line), 600 (dashed line), and 1000 $M_{\odot}$ (dot-dashed line).
The horizontal dotted line indicates the present-day
mass density in black holes in the local universe estimated by
Merritt \& Ferrarese (2001).
\label{plot2}}
\end{inlinefigure}

\section{Implications}

\subsection{The End of the Population~III Era}

One may naively expect that Population~III star formation
could continue
down to moderate redshifts, $z\sim 5-10$.
We argue, based on the results of our model, that
either the completion of reionization or prompt enrichment
to a level of $\bar{Z}\sim 10^{-4}Z_{\odot}$ by $z\sim 15$ should mark the
end of the Population III era.

When a large fraction of the IGM is reionized,
effective photo-evaporation
of minihalos by internal and external sources occurs
(e.g., Shapiro, Iliev, \& Raga 2003), and virtually no gas can be
trapped within low mass halos.
Furthermore, SN explosions cause nearly complete evacuation of
the host halos (see Paper~I). Therefore,
regardless of the efficiency of the proposed positive feedback
on molecular hydrogen formation
(e.g., Ferrara 1998; Oh 2001), it should take some time for the heated and expelled
gas to cool and {\it condense}.
Subsequent star formation can take place only after the gas falls back into
the halos again. Accordingly, there should be an appreciable time gap
between the first star formation and the onset of the second-generation
star formation.
We conservatively estimate this `recuperation time' (Sokasian et al. 2003a)
to be of the order of the dynamical time scale
for the halos. The dynamical time scale is given by
\begin{equation}
t_{\rm dyn}=\sqrt{\frac{3\pi}{32G\rho}}\mbox{\ ,}
\end{equation}
where $G$ is the gravitational constant and $\rho$ the
total mass density of a halo.
Assuming that the mass density of a halo (including dark matter)
is 200 times the cosmic mean density,
we compute the dynamical time for halos collapsing at $z=15$ to be $\sim 50$ Myr.
Since the cosmic time difference between $z=15-18$ (or $z=13-15$)
is $\sim 60$ Myr for the standard $\Lambda$CDM model, star formation is
likely to be quenched for at least a duration of
$\Delta z\sim$ a few in redshift.
It might seem that halos with virial temperature greater than
$10^4$~K are unaffected by photo-heating due to reionization
(e.g., Oh \& Haiman 2002).
We point out that, in the hierarchical universe, {\it all}
the massive halos used to be minihalos at earlier epochs. Hence,
according to our
model of Population III star-formation, the gas within them is
likely to have been
already expelled either by radiation from the Population III stars or by SN
explosions.
The situation would be slightly different in halos
that are rapidly growing in mass because such halos may be
able to trap the ionized hot gas. However, SN explosions with energies
$\sim 10^{53}$ ergs are still able to cause an outward gas flow
even in the large halos.
If the feedback from the first massive star on its vicinity
occurs by z=15, as in our model,
then the formation of the second-generation stars
in the same region (i.e., protogalaxy) will begin at $z \lesssim 13$.
Further,
very massive stars are already pulsationally
unstable when the metallicity is $\sim 10^{-4} Z_{\odot}$,
even if they are formed (Baraffe et al. 2001).
The minimum metallicity floor is established in all the star-forming regions
by $z\sim 15$ if the first stars explode as pair-instability SNe,
thus causing a `global' regulation of the star formation activity
lasting until $z\sim 13$.
%
Note that significant cosmic variance may be induced by large-scale 
($\gg 1$Mpc) density fluctuations. The cosmic scatter in the 
number density of the star-forming clouds should then cause 
a substantial spread in the epoch of self-termination
(e.g. Scannapieco, Schneider \& Ferrara 2003). 

An alternative scenario for second-generation star formation
is self-propagating star formation in the thin, dense shell formed by early
SN explosions (e.g., Ostriker \& Cowie 1981; Wandel 1985;
McCray \& Kafatos 1987).
Such phenomena are indeed observed in the local universe
(e.g., Dopita et al. 1985;
Efremov \& Elmegreen 1998) and may have taken place
in the early universe as well (Mackey et al. 2003;
Salvaterra, Ferrara, \& Schneider 2003).
For the particular case we simulated in Paper I,
we found that no gas clumps are formed in the dense shell which
are Jeans-unstable. This may, however, be because of the 
still insufficient mass resolution of the simulation.
In order to properly address the question of the shell 
fragmentation and the formation of proto-stellar clouds,
simulations with an extremely
high mass resolution are needed. 
Also, more complex configurations, 
such as collisions between neighboring
blast waves originating in more massive systems,
may result in the formation of gravitationally bound fragments.
We pursue this possibility further in a forthcoming work
(Bromm et al., in preparation).

\subsection{Thomson Optical Depth and Mean Metallicity}

In \S~2, we have shown that the production of ionizing photons and
heavy elements by massive Population III stars is described by fundamental
constants in the case where these stars explode as pair-instability
supernovae.
Here, we derive
an approximate but concise
relationship between the evolution of the mean metallicity and the total Thomson
optical depth, $\tau_{e}$, as follows.

The differential optical depth to Thomson scattering corresponding to
the redshift interval $dz$ is
\begin{equation}
d\tau_{e}=\sigma_{\rm T}\,n_{e}(z)\,c\frac{dt}{dz}\,dz,
\end{equation}
where $\sigma_{\rm T}=6.65\times 10^{-25} {\rm cm}^2$ is the Thomson scattering
cross section, and $n_{e}(z)$ the mean electron number density at $z$.
The mean electron number density is
given by
\begin{equation}
n_{e}(z)=n_{e}(0)(1+z)^{3} x_{e}(z),
\end{equation}
where $x_{e}$ is the ionization fraction at $z$, and $n_{e}(0)$
the present-day number density of electrons.
We obtain the contribution from an interval between $z_{1}$ and $z_{2}$ ($z_{1} < z_{2}$),
\begin{eqnarray}
d\tau_{e}|_{z_1\rightarrow z_2}&=&\sigma_{\rm T}\,n_{e}(0)\,x_{e}(z)
\frac{c}{H_{0}}\\ \nonumber
&\times& \frac{(1+z)^2}{(\Omega_{\Lambda}+\Omega_{\rm m}(1+z)^3)^{1/2}}\, dz\\ \nonumber
&\approx& \sigma_{\rm T}\,n_{e}(0)\,x_{e}(z)
\frac{c}{H_{0}}\left(\frac{1+z}{\Omega_{\rm m}}\right)^{1/2}
\end{eqnarray}
We have assumed a flat $\Lambda$CDM model with
$H(z)=H_0[\Omega_{\Lambda}+\Omega_{\rm m}(1+z)^3]^{1/2}$ and
the last approximation holds at high redshifts, certainly for $z>10$.

Neglecting recombination, we can approximate, to the lowest order,
the ionization fraction as
\begin{equation}
x_{e}(z)\propto q(z).
\end{equation}
{}From equation (\ref{eq:zmean}) we obtain for pair-instability SNe,
\begin{equation}
x_{e}(z) \sim \frac{\bar{Z}(z)}{\zeta_{\rm PISN} f_{\rm PISN}}\mbox{\ .}
\end{equation}
Absorbing all the constant factors into
a single quantity $\Upsilon$, including $f_{\rm PISN}$,
we express the concise relation as
\begin{equation}
\int d\tau_{e}\simeq\Upsilon \int_{z} \bar{Z} (z) (1+z)^{1/2} dz.
\label{eq:tau_Z}
\end{equation}
Thus, to the lowest order, the evolution of the ionization fraction should be
accompanied by the increasing metallicity following equation (\ref{eq:tau_Z}).
Integrating equation (\ref{eq:tau_Z}) to $z=z_{\rm chem}$, when $\bar{Z}=5\times 10^{-4}Z_{\odot}$,
we obtain
\begin{equation}
\Delta \tau_{\rm e, PopIII} \sim 0.06.
\end{equation}
This is the
contribution from early Population III stars expected
in our model.

\subsection{Metallicity in the Lyman-$\alpha$ Forest at $z \sim 3-5$}

Metals produced by the early Population III stars should exist ubiquitously
in various systems at low redshift.
Miralda-Escud\'{e} \& Rees (1997) argue that the observed metallicity
in the Lyman-$\alpha$ forest at $z\sim 3$ is consistent with a simple
picture
in which the metals originate from an early generation of stars.
The average metallicity $\bar{Z}=10^{-2}Z_{\odot}$ in the Lyman-$\alpha$ forest
at $z\sim 3$ is still much higher than the mean metallicity achieved
at the end of the Population~ III era, $z\sim 15$, whereas
Songaila (2001) argues that a minimum universal metallicity of about
$3.5\times 10^{-4}Z_{\odot}$ is found in $z\sim 5$ Lyman-$\alpha$ clouds.
Remarkably, this is very close to $\bar{Z}\sim 3\times 10^{-4}Z_{\odot}$
(Figure 1) achieved by the early massive Population III stars.
Since the Population III star formation is self-regulating
owing to the global radiative feedback and metal enrichment,
this value is expected to be approximately the maximum which early Population III stars
alone can achieve.
Songaila (2001) estimates the mass density of {\ion{C}{4}} between $z=1.5$
and $z=5$ to be $\Omega_{\rm C_{\rm IV}, IGM} = (2.5-7) \times 10^{-8}$.
Assuming  {\ion{C}{4}} is the dominant ionization stage with ionization fraction of 0.5
(Songaila 1997), the mass density in carbon is $\sim (5-14) \times 10^{-8}$.
For a very massive zero-metalicity star, the carbon yield is
$\sim 4 M_{\odot}$ and nearly constant for the relevant core mass
(Heger \& Woosley 2002). From this we obtain the mass density
in carbon (more precisely, $^{12}$C) for our model with $f_{\rm PISN}=1$,
\begin{equation}
\Omega_{\rm C_{\rm IV}, PopIII} \approx 10^{-8}
\end{equation}
at $z=15$. Although smaller than the value that Songaila (2001) obtains
for the diffuse IGM at $z\sim 5$, this is not negligible.
Indeed, a fraction of the carbon, up to 20\%, in the IGM may originate from
the early generation Population III stars.

At $z\sim 3$, the bulk of the metals found in
the Lyman-$\alpha$ forest may also come from ordinary stellar
populations formed in galaxies at lower redshift, $z<5$
(e.g., Adelberger et al. 2003).
Interestingly, the metallicity evolution computed by Hernquist \& Springel (2003)
based on their analytic model of star-formation in galaxies
predicts that the average metallicity reaches $\sim 10^{-2}Z_{\odot}$
at $z\sim 3-5$, in good agreement with the observed value in the Lyman-$\alpha$ forest.

\subsection{Enrichment of the Intracluster Medium}
X-ray observations of galaxy clusters reveal that the ICM
is enriched to a significant fraction of the solar metallicity
(e.g., Mushotzky et al. 1996; White 2000).
In fact, a large fraction ($\sim 75$\%) of a typical cluster's
endowment with heavy elements exists in the ICM
rather than in cluster galaxies (Renzini 1997).
Clusters retain all the metals produced in them, and the determination of
the abundances for several elements in the ICM is relatively
straightforward, without involving complicated modeling.
Thus, the elemental abundances in the ICM offer
a clean template against which any theoretical models can be tested.
Loewenstein (2001) suggests that pre-enrichment by a generation of massive
Population III stars may account for the abundance anomalies of oxygen,
silicon and iron, observed
in the ICM.
The recent analysis of the relative overabundance of Si with respect to Fe
in the ICM also indicates the
possible contribution from massive Population III stars.
Baumgartner et al. (2003) argue that the observed relative overabundance
of Si can be explained by Population III progenitors with a 70 $M_{\odot}$ He core.
While including the contribution from massive Population III stars
is attractive, the required abundance appears very large.
Baumgartner et al. (2003) estimate that the total mass
in Population III stars should be at least
$\Omega_{\rm PopIII} \sim 10^{-5}$, in units of the cosmic density parameter,
whereas our model predicts only
\begin{equation}
\Omega_{\rm PopIII} \sim 5\times 10^{-7},
\end{equation}
at $z=15$. Baumgartner et al. (2003) argue that primordial star formation is enhanced
in overdense regions which later become clusters.
However, while this might seem plausible, the strong self-regulation
of early star formation primarily tends to shift star formation to
earlier epochs in overdense
regions, rather than
locally enhancing the total amount of formed stars.
We therefore conclude that metals originating from the earliest generation stars
cannot be responsible for the observed abundance anomalies in the ICM.

\subsection{Dust Formation and H$_{2}$ Production}
We have computed the mean metallicity of the IGM assuming implicitly
that all the metals produced and ejected by SN explosions
will remain in the gas phase. This is not strictly true; a fraction of the metals
will be in the form of dust and hence do not directly affect the gas chemical properties
(see however the effect of promoted molecular hydrogen formation discussed later in this
section). Schneider, Ferrara, \& Salvaterra (2003a) and Nozawa et al. (2003)
calculate the dust depletion factor
for pair-instability SNe to be up to 30-40\%.
Nozawa et al. (2003) also conclude that the dust
formation rate and the relative abundances of the dust grains of various compounds
depend sensitively on the mass of the progenitor and on
the mixing efficiency in
 the ejecta. Another complexity is that the produced dust grains are
destroyed by
thermal sputtering with ions and electrons, and by grain-grain collisions
in post-shock regions (Spitzer 1978).
Most important, the reverse shock
formed in the early stages of a SN
explosion can destroy a significant fraction of the dust grains 
produced in the
ejecta (McKee 1989).
Unfortunately,
the efficiency of dust destruction by shocks is not well known
and we need to resort to a simple parametrization.
If, for example, the destruction efficiency is very low and nearly all the formed dust
grains survive, are dispersed and remain in the IGM, the cumulative amount of dust could be
significant. We choose the dust depletion factor to be $f_{\rm dpl}$=0.4 (Schneider et al. 2003a),
and assume a low destruction efficiency, $e_{\rm dest}$=0.1.
Using the global formation rate for Population III stars
in equation (\ref{eq:sfr_popIII}) and assuming $f_{\rm PISN}=1$,
we calculate a conservative upper limit
for the mass density of dust in the IGM,
\begin{equation}
\Omega^{\rm IGM}_{\rm dust} \approx 10^{-7}
\end{equation}
by $z=15$. This is much smaller than, for example,
the density of intergalactic dust,
$\Omega^{\rm IGM}_{\rm dust} \sim 10^{-4-5}$ at $z=0.5$
(Aguirre 1999; see also Pei, Fall, \& Hauser 1995; Croft et al. 2000),
and thus appears to make negligible contribution to the dust in the
diffuse IGM even if the grains survive until the present epoch.
We point out, however, that even a small quantity of dust in the gas
accelerates H$_{2}$ production
and hence affects the efficiency of gas cooling and star formation
when the gas metallicity is still low (Hirashita et al. 2002).
Todini \& Ferrara (2001) argue that H$_{2}$ production on dust grains dominates
over the usual H$^{-}$ channel when the dust-to-gas ratio
is greater than 5 percent of the local Galactic value $D_{g}\sim 6\times 10^{-3}$.
Our model predicts that the mean dust-to-gas ratio is $\sim 3\times 10^{-6}$ at $z=15$
for $f_{\rm dpl}$=0.4 and $e_{\rm dest}$=0.1,
still an order of magnitude smaller than the critical value.
H$_{2}$ production on dust grains could
still be important if the dust-to-gas ratio were locally high,
either because the distribution is inhomogeneous, or because
the dust grains remain preferentially within a confined region around the
SN site.

\subsection{Building-up an Early X-ray Background}

It has been suggested that X-rays could act as a source of positive feedback
by boosting
the free electron fraction in distant regions. This would in turn
promote the formation of H$_{2}$, hence enabling enhanced cooling of
primordial gas
(e.g., Haiman, Rees, \& Loeb 1996; Oh 2001; Glover \& Brand 2003).
Apart from early quasars (e.g., Eisenstein \& Loeb 1995; Bromm \& Loeb 2003;
 Madau et al. 2003),
SNRs are the only plausible X-ray sources
at very high redshifts.
High redshift SNRs cool via
inverse-Compton cooling, losing a large fraction of their energy to
CMB photons (Oh, Cooray, \& Kamionkowski 2003).
Some fraction of the thermal energy is also converted into X-rays via thermal
bremsstrahlung, and yet another unknown fraction goes 
into relativistic electrons
which could emit X-rays via inverse-Compton scattering of CMB photons.
Unfortunately, the relative 
fractions of energy delivered to
these processes is highly uncertain, and likely to be time-dependent
during the evolution of the SNRs.
In order to obtain a rough order-of-magnitude estimate, we calculate the evolution
of an early X-ray background using some working assumptions as follows.
We consider only X-rays with energies greater than 1 keV, because
the IGM is nearly optically thin to these X-rays,
and assume further that a fraction $c_{\rm X}$
of the total SN explosion energy is converted to radiation in the relevant X-ray band,
between 1 keV and 5 keV.
We set, somewhat arbitrarily, $c_{\rm X}=0.1$ as our fiducial value.
Note that this value is significantly larger than
the one assumed by Glover \& Brand (2003) as a {\it maximum} efficiency case,
based on the estimate of Terlevich et al. (1992) for X-ray bright remnants
in dense environments, $c_{\rm X} \approx 0.01$.
Hence, the results presented below provide a conservative upper limit.

For the emissivity of the SNRs, we rather crudely model it to be
described by a power law $L_{\rm \nu}\propto \nu^{-1}$
in the narrow (1-5 keV) energy range.
By noting that radiative cooling in the SNRs occurs
mainly in the first few million years for
the original SN energy of $\sim 10^{53}$ ergs (see section 2.3),
the time-averaged luminosity in X-rays from a single SNR is
estimated to be
\begin{equation}
L_{\nu} \approx 4.2\times 10^{20} \left(\frac{\nu_{0}}{\nu}\right)^{-1}  \;{\rm ergs\; s}^{-1} {\rm Hz}^{-1},
\label{eq:L_nu}
\end{equation}
where $h\nu_{0}=1$keV.
We compute the evolution of the diffuse X-ray flux by solving the
cosmological radiative transfer equation (e.g., Peebles 1993)
\begin{equation}
\left(\frac{\partial}{\partial t}-\nu H(z)\frac{\partial}{\partial \nu}\right) J
= -3H(z)J+\frac{c}{4\pi}\epsilon\mbox{\ ,}
\end{equation}
where $J$ is the specific intensity in units of ergs s$^{-1}$ cm$^{-2}$ Hz$^{-1}$ sr$^{-1}$,
$H(z)$ the Hubble parameter at redshift $z$, $c$ the speed of light,
and $\epsilon$ the proper volume-averaged emissivity.
We use this unconventional unit for the X-ray flux in order to
facilitate comparisons with the commonly considered soft-UV background radiation.
We compute the volume emissivity by combining equation (\ref{eq:L_nu})
with our model SFR (equation [13], see also Figure 1),
assuming that all the Population III stars die as pair-instability SNe
($f_{\rm PISN}=1$). This simple model again serves as a maximal case
in terms of X-ray emission efficiency.

Figure \ref{fig:xray} shows the evolution of the X-ray background radiation intensity
at 1 keV. It is clear that the flux reaches only up to $10^{-24}$ ergs s$^{-1}$ cm$^{-2}$ Hz$^{-1}$ sr$^{-1}$
by $z=15$.
Machacek, Bryan, \& Abel (2003) performed numerical simulations of pre-galactic
object formation to examine the effect of the positive feedback from a
diffuse X-ray background
in the early universe. They conclude that the net effect is quite small
for X-ray fluxes below $\sim 10^{-23}$ ergs s$^{-1}$ cm$^{-2}$ Hz$^{-1}$ sr$^{-1}$.
 To result in a positive feedback that counteracts the negative feedback
due to photo-dissociating radiation, an order of magnitude higher flux
is necessary, which is two orders of magnitude higher
than our estimate for the X-ray flux at $z=15$. Overall,
the effect of the early X-ray background caused by SNRs
to the primordial gas cooling appears unimportant at $z>15$.

\subsection{The G-dwarf Problem}
It has often been suggested that the
paucity of metal-poor stars in the solar neighborhood
relative to the prediction of theoretical models,
the so-called G-dwarf problem, can be resolved
by invoking early massive Population~III stars (Cayrel 1986; see also Larson 1998
for a recent review).
In fact, the kind of pre-enrichment by early generation
stars proposed here may offer an excellent resolution
if the metallicity of the gas is sufficiently high initially.
The simplest models require a significant degree of pre-enrichment
in the Galactic disk (e.g., Binney \& Tremaine 1987), $Z_{\rm init}=0.25Z_{\odot}$, whereas
our model predicts far too  small a metallicity accomplished by Population III
stars alone, $\sim 10^{-4}Z_{\odot}$.
Thus, considering the self-regulating 
nature of early star formation (see \S3.1),
we conclude that massive Population III stars 
do little to resolve the classical G-dwarf problem.

\begin{inlinefigure}
\resizebox{8.8cm}{!}{\includegraphics{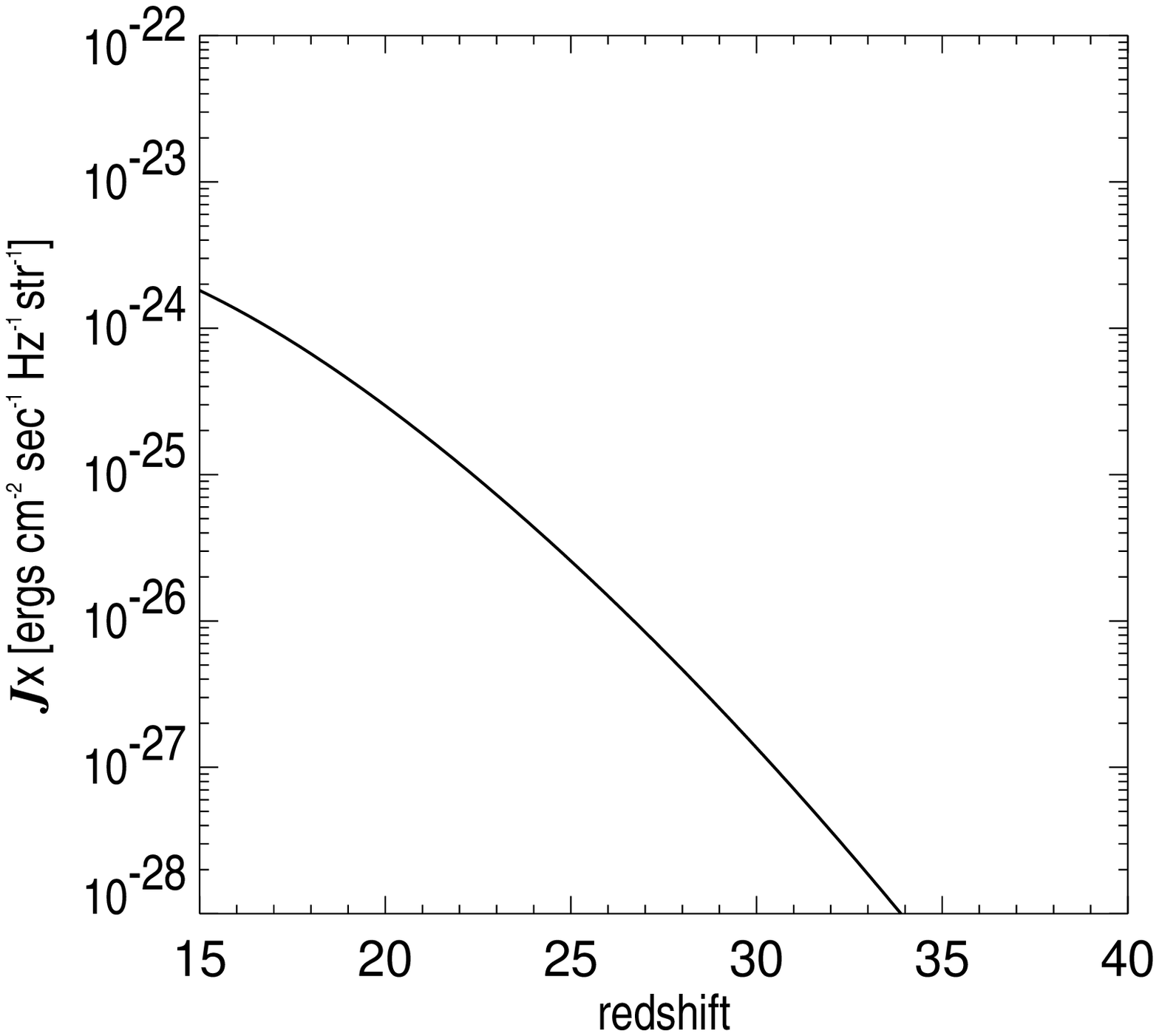}}
\caption{The evolution of the X-ray background radiation intensity.
We use unconventional units,
ergs cm$^2$ s$^{-1}$ Hz$^{-1}$ sr$^{-1}$,
in order to compare  with the typical intensity of the soft-UV
background radiation $J_{\rm UV} \sim 10^{-21-23}$ (in the above units),
which is responsible for the negative feedback on H$_{2}$ cooling.
\label{fig:xray}}
\end{inlinefigure}

However, massive Population III stars
may account for a modern version of this problem
-- the paucity of extremely metal-poor stars in the Galactic halo.
As we have shown in previous sections,
a prompt enrichment to $\sim 10^{-4}Z_{\odot}$ is achieved
by $z=15$. In fact, the gas metallicity around the SN sites,
where most plausibly the second generation stars will be formed,
is higher than this {\it mean} value, and thus, within our model,
the second generation stars are expected to have metallicities
at or above this critical value.
Observationally, with the important exception of the ultra-metal poor
star discovered by Christlieb et al. (2002),
the paucity of stars with [Fe/H] $< -4$ is apparent
in our Galaxy. It would be extremely important
to derive the abundance pattern  of such stars 
through a systematic search (e.g.,
Christlieb 2003).

\subsection{Running Spectral Index Models}

The rapid metal-enrichment by massive Population III stars
has important implications for some variant CDM models.
Based on a joint analysis of the first-year {\it WMAP} data,
the 2dF galaxy redshift survey, and the Lyman-$\alpha$ forest,
Spergel et al. (2003) conclude that cosmological models with a
running primordial power spectrum are favored
over power-law models with $n\approx 1$ (see also Peiris et al. 2003).
However, post-{\it WMAP} studies indicate that
the observed high optical depth is not achieved in such models with
a strong running, because the formation of pre-galactic size
objects is systematically delayed owing to the suppressed
power on small scales (e.g., Somerville et al. 2003).
Chiu, Fan, \& Ostriker (2003) and Avelino \& Liddle (2003) indeed
showed that strong running of the primordial power
results in considerable delay of reionization.
Yoshida et al. (2003b) carried out cosmological hydrodynamic
simulations of early structure formation and concluded
that the running spectral index (RSI) model is incompatible with the
observed high Thomson optical depth.
However, Yoshida et al. also argue that the resulting
Thomson optical depth can be made marginally consistent with the {\it WMAP}
data, {\it if} the ionizing photon production
rate is extremely high in galaxies at $z>6$.
This practically indicates that, in the RSI model,
a top-heavy IMF for the stars formed in galaxies must be maintained
over a long period, $6<z<20$.

The problematic nature of this requirement can be appreciated
by noting that the total elapsed time to $z=6$ in a
$\Lambda$CDM cosmology is approximately $10^9$ years.  Very massive
stars, like those considered here, have lifetimes $\sim 2\times
10^6$ years.  Thus, if star formation commences at high
redshifts, $z\simgt 20$, nearly 500 generations of these
high-mass stars could have formed.  As we indicate below, the
metals from these stars must be ``hidden'' to an
improbably high degree of
precision for Population III star formation not to
self-terminate well before $z=6$.

Here, we show that, in scenarios such as the running spectral
index model,
a large quantity of metals is necessarily produced
if it is assumed that Population III star formation continues
until $z=6$.
Let us consider a simplified model in which the stellar IMF is
essentially a $\delta$-function at $300 M_{\odot}$, i.e.,
all the stellar baryons are converted into very massive stars.
This is the most efficient way of getting ionizing photons
out of a certain mass of baryons converted into stars,
because the luminosity per stellar mass is essentially a constant
for metal-free stars with mass exceeding $300 M_{\odot}$.
In this case we may simply assume a high ionizing photon production rate
per stellar mass of $1.6\times 10^{48}/M_{\odot}$.
As in Yoshida et al. (2003b), the star-formation rate density for the
RSI model is computed using the same functional fit (equation [\ref{eq:sfr_popIII}])
with $p=1.5$, $\tilde{q}=3.05$, $r=1.85$, $\tilde{\alpha}=0.01$, $\beta=0.057$, and
$\dot{\rho}_{*}(0)=0.0176 M_\odot$ yr$^{-1}$ Mpc$^{-3}$.

\begin{inlinefigure}
\resizebox{9cm}{!}{\includegraphics{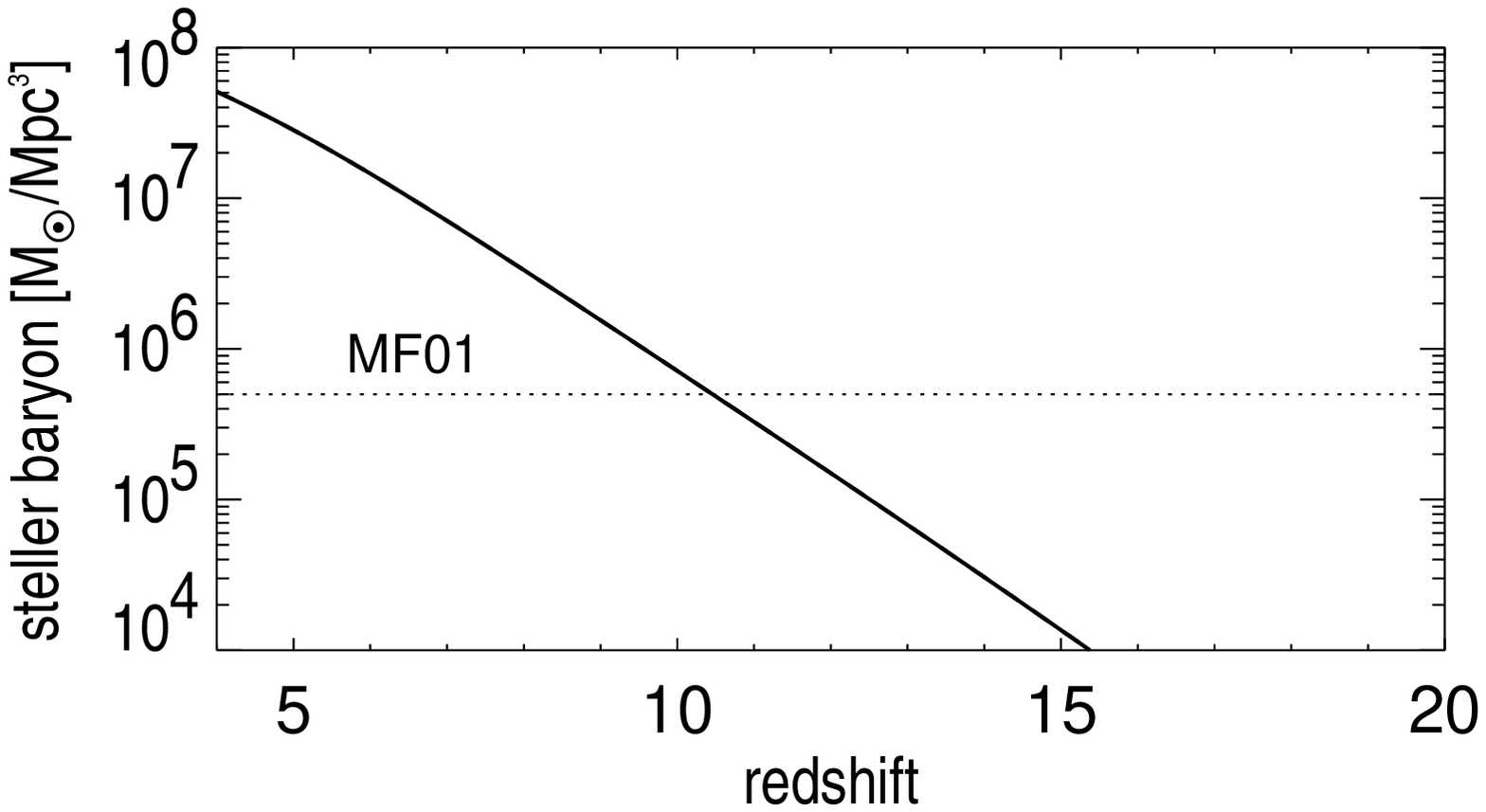}}
\vspace{-3mm}
\resizebox{9cm}{!}{\includegraphics{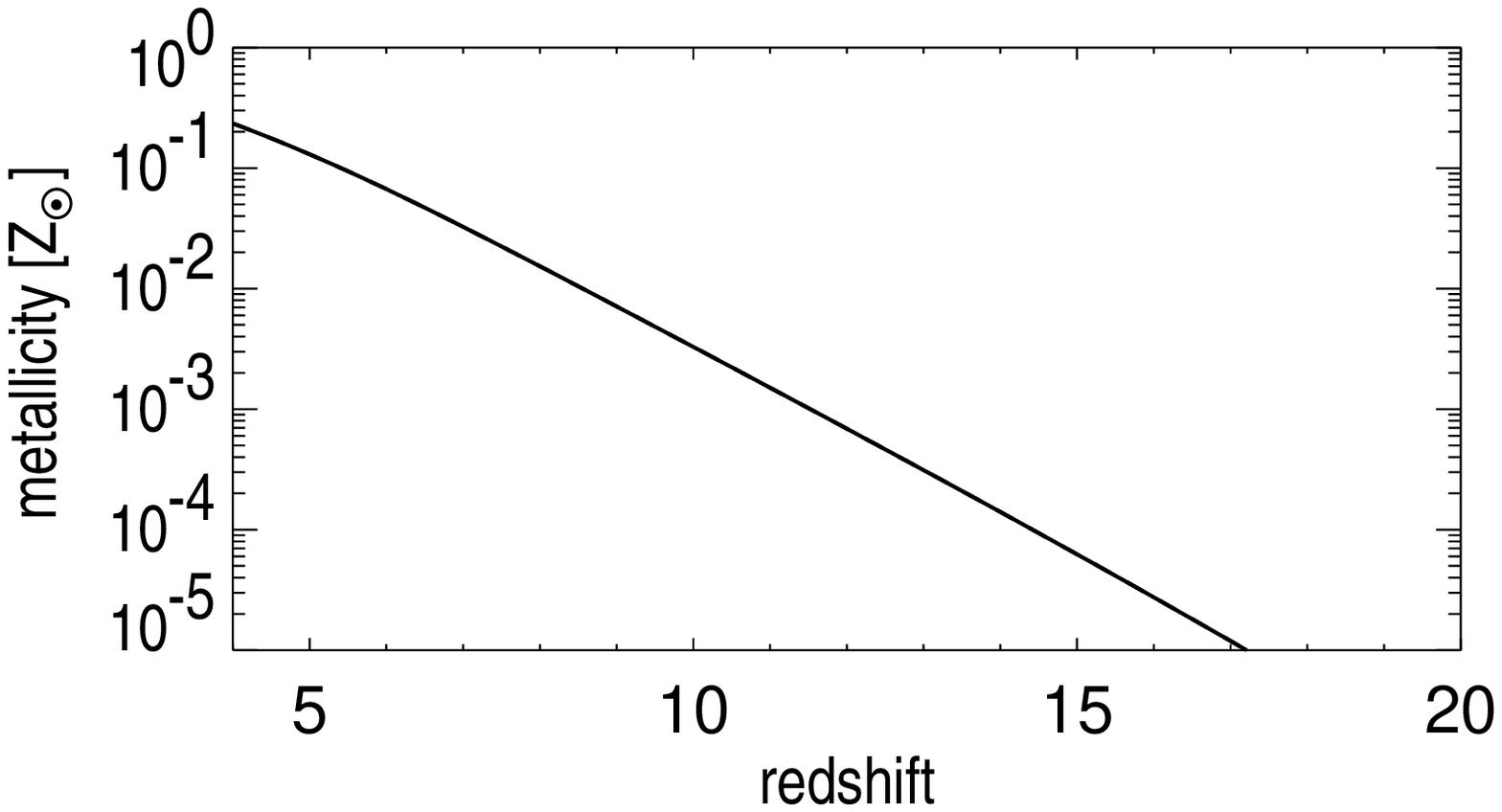}}
\caption{We plot the cumulative mass of stellar baryons, which is
locked up in black holes in our extreme model (top),
and the mass-weighted mean metallicity (bottom) for the RSI model
on the assumption that all the stellar baryons are processed through
very massive stars.
\label{fig:sfr_rsi}}
\end{inlinefigure}

Figure \ref{fig:sfr_rsi} shows the cumulative mass density
of stellar baryons. It directly corresponds to
the mass density of black holes if we assume that {\it all} the stars
become BHs, i.e., their masses are greater than $\sim 300 M_{\odot}$.
In the plot, the dotted line indicates the present-day
mass density in black holes in our local universe estimated by
Merritt \& Ferrarese (2001). Clearly the mass density increases beyond
the {\it present-day} mass density by $z=10$.
On the other hand, if all the stars are massive, but within the pair-instability
SN range of $140 M_{\odot} < M < 260 M_{\odot}$, up to half of the stellar mass
can be processed into heavy elements and expelled.
Clearly these models are idealized, but
they allow us to explore the possible parameter space.
On these assumptions and using the SFR for the RSI model,
we compute the mass-weighted mean metallicity.
The bottom panel of Figure \ref{fig:sfr_rsi} shows the evolution of the mean
metallicity $\bar{Z}$. Remarkably, the critical metallicity $\sim 10^{-4} Z_{\odot}$
is achieved already by $z\sim 14$ and it reaches $\sim 10^{-1} Z_{\odot}$
by $z=5$.
It is now clear that, if we resort to the extremely top-heavy IMF in the RSI model
to produce enough UV photons to reionize the universe early on, the total amount of heavy
elements quickly rises up to a point where the gas chemical properties are expected
to change (Bromm et al. 2001b). It is conceivable
that all the massive stars collapse to black holes so that the IGM
remains chemically pristine without metal-pollution, but then perhaps too much
baryonic mass goes into black holes, which appears to be incompatible with
the recent estimate of the local mass density in black holes.
In summary, a running of the primordial power spectrum
seems to be incompatible with various observations in one way or another.
A similar conclusion 
is expected to obtain for other variant CDM models, such as
those involving warm dark matter (e.g., Somerville et al. 2003;
Yoshida et al. 2003c).


\section{Summary}

We have revisited the cosmological consequences of an early generation of Population~III stars.
We have primarily considered very massive ($\sim 100-300 M_{\odot}$) stars and
used a
detailed model of the global star formation rate which is derived from
semi-analytic modeling
and the results of cosmological hydrodynamic simulations
that include radiative feedback effects.
We have shown that the first stars alone can produce a sufficient
number of photons to reionize a significant volume of the IGM by $z\sim 15$,
if the stars are very massive,
in agreement with cosmological radiative transfer calculations
(e.g., Sokasian et al. 2003a).
Furthermore, on the assumption that the majority of the stars
explode as pair-instability SNe, the mean IGM metallicity
increases to the critical value above which the gas cooling efficiency is
greatly enhanced. This may lead to the formation of ordinary stellar populations including
low-mass stars (e.g. Mackey et al. 2003).
We also discussed possible implications of the results.
Although a broad range of observations indicates that the kind of massive stars
we considered in the present paper existed in the early universe,
the simple model may not be able to explain the observations in details.
Clearly, further studies using more sophisticated models
are necessary to examine the cosmological consequences of
early star formation. In particular, we will explore the contribution
from the second (and subsequent) generation stars forming at $z<13$
in future work.

The evidence for an early generation of very massive stars may well be
strengthened by future observations.
The ongoing operation of {\it WMAP} will yield a more precise
value for the total optical depth to reionization.
In the longer term, post-{\it WMAP} CMB polarization experiments such
as {\it Planck} will probe the reionization history
(e.g., Kaplinghat et al. 2003). Detection of the second-order
polarization anisotropies on arcminute scales can place
strong constraint on the details of reionization (e.g., Liu et al. 2001).
Near-infrared observations of afterglows from high-redshift
gamma-ray bursts can also be used to probe the reionization history
at possibly $z>10$ (Barkana \& Loeb 2003; Inoue et al. 2003;
Yoshida \& Bloom 2003, in preparation).
Mapping the morphological evolution
of reionization may be possible by observations of
redshifted 21cm emission by the Square Kilometer Array
and LOFAR (e.g., Tozzi et al. 2000; Furlanetto, Sokasian, \& Hernquist 2004).
Data from these future observations will provide a more
complete picture of cosmic reionization
and will enable us to distinguish the sources responsible for
reionization.
The precise measurement of the near-IR cosmic background radiation
will constrain
the total amount of light from early generation
stars (Santos, Bromm, \& Kamionkowski 2002; Salvaterra \& Ferrara 2003).
Ultimately, direct imaging and spectroscopic observations of high redshift
star clusters by the {\it James Webb Space Telescope} will
probe the evolution of stellar populations up to $z\sim 10-15$
(e.g., Stiavelli, Fall, \& Panagia 2003; Tumlinson, Shull, \& Venkatesan 2003).

Finally, measurements of the relative abundances of
various heavy elements in metal-poor stars
should provide valuable information on the chemical history
of the universe
(e.g., Burris et al. 2000; Norris et al. 2002).
Interestingly, a strong argument against
very massive ($>140 M_{\odot}$) stars comes from
the observed abundance pattern of C-rich, extremely Fe-deficient
stars (Christlieb et al. 2002; Umeda \& Nomoto 2003;
but see Schneider et al. 2003b).
It remains to be seen whether or not such stars are
truly second generation stars and their elemental abundances
should precisely reflect the metal-yield from
the first SNe. Observations of a large number of extremely
metal-poor stars will construct better statistics
(Norris, Ryan, \& Beers 2001) and improve
constraints on any models for the early chemical evolution.
Understanding the origin of the first heavy elements
in the universe and the nature of the sources
that are responsible for cosmic reionization will require the concerted use of
data from these broad classes of observations.

\acknowledgments
We thank Evan Scannapieco for helpful comments,
and Aaron Sokasian for providing us with the SFR data.
NY acknowledges support from the Japan Society of Promotion of Science
Special Research Fellowship (grant 02674).
This work was supported in part by NSF grants ACI
96-19019, AST 98-02568, AST 99-00877, and AST 00-71019 and NASA ATP
grant NAG5-12140.

\end{document}